# A compact single channel interferometer to study vortex beam propagation

**Sruthy J. Lathika[1,2], A. Vijayakumar[1,2,*], and Shanti Bhattacharya[1,**]**

[1]Department of Electrical Engineering, Indian Institute of Technology Madras, Chennai – 600036, India.
[2]The authors contributed equally to the work.
[*]Currently at Centre for Micro-Photonics, Faculty of Science, Engineering and Technology, Swinburne University of Technology, Hawthorn VIC 3122, Australia.

**Corresponding author: shantib@iitm.ac.in

## ABSTRACT

We propose and demonstrate a single channel interferometer that can be used to study how vortex beams propagate through a scatterer. The interferometer consists of a multifunctional diffractive optical element (MDOE) synthesized by the spatial random multiplexing of a Fresnel zone plate and a spiral Fresnel zone plate with different focal lengths. The MDOE generates two co-propagating beams, such that only the beam carrying orbital angular momentum is modulated by an annular stack of thin scatterers located at the focal plane of the Fresnel zone plate, while the other beam passes through the centre of the annulus without any modulation. The interference pattern is recorded at the focal plane of the spiral Fresnel zone plate. The scattering of vortex beams through stacks consisting of different number of thin scatterers was studied using the proposed optical setup. Conflicting results have been reported earlier on whether higher or lower charge beams suffer more deterioration. The proposed interferometer provides a relatively simple and compact means of experimentally studying propagation of vortex beams through scattering medium.

## 1. Introduction

Light propagation through scattering and turbid medium produce speckle signatures from which it is often difficult to extract useful information. Different techniques have been developed in order to look through both static as well as dynamic scatterers [1-5]. These techniques can be broadly classified into two types. In the first type, optical signal processing [6-8], correlation techniques [9-13] and phase retrieval algorithms [14,15] have been developed to decode a speckle signature and convert it into useful information. A second direction of research is to develop adaptive aberration correction methods to compensate for the disturbance introduced by the scattering medium [16, 17]. In the latter approach, structured light beams with scatter-resistant characteristics are used to reduce the aberrations introduced by the scatterer [18, 19].

Vortex or orbital angular momemtum (OAM) beams have a unique complex amplitude, which makes them attractive for various applications. For example, Laguerre-Gaussian (LG) beams have been used in optical trapping [20] and optical communication [21] experiments. Besides, LG beams have also shown scattering-resistant properties in various studies [22-25] and several techniques have been developed to aid probing and communicating through scattering medium using LG beams [26, 27]. In most of these techniques [25, 28], the deterioration of the phase of the beam is studied using an interferometer. With the need for interferometry, the footprint of the experiment increases, as more components and vibration isolation, have to be added to the experiment. There are few interference-less optical configurations and techniques used to measure the level of deterioration of the LG beam [24, 26] but the information is incomplete without measurement of phase.

In this manuscript, we propose a single channel optical configuration based on a multifunctional diffractive optical element (MDOE) for studying the robustness of LG beams of different charges versus level of scattering. The advanced optical configuration does not require a vibration isolation system, beam splitting, etc., and therefore, is extremely compact. Secondly, the interferogram is designed to directly show whether the charge of the beam changed during propagation, in terms of the number of radial intensity lobes. In that sense, the measurement of the deterioration of LG beams through thin scatterers can be considered a pattern recognition problem.

Several groups have reported theoretical and simulation studies of OAM beam propagation in a scattering medium. Research work by Ou's and Wang's groups [29, 30] show that the scattering intensity decreases as topological charge increases. This behaviour can be accounted to the large diameters of the dark central vortex region of OAM beams with higher topological charges resulting in lesser interaction with the particles in the turbid medium. On the other hand, the work under Qu [31] shows that the scattering intensity increases with an increase in the topological charge. There is also a report [25] stating that the performance of OAM beams is better in scattering media than Gaussian beams, when tightly focused. On the other hand, their studies showed that Gaussian beams fared better than OAM beams when the beams were not that tightly focused. While the results depend on a number of factors including the type of scattering media under study, the characteristics of the OAM beam propagated, and the performance parameters measured; it is clear that more detailed studies are required to resolve the many different results presented so far. The compact interferometer is proposed as an elegant way to carry out further studies in this important field.

## 2. Methods

The optical configuration of the single channel, dual beam interferometer is shown in Fig. 1(a) and the synthesis of the MDOE, with phase $\phi_{MDOE}(x, y)$, is shown in Fig. 1(b). The MDOE is synthesized by random multiplexing of two diffractive functions, namely a Fresnel zone plate [32] and a spiral Fresnel zone plate [33, 34] with different focal lengths. The phase of the Fresnel zone plate ($FZP$) is given by $\Phi_{FZP}(f) = \{-\pi/\lambda f\}R^2$, where $R$ is the radial coordinate and the focal length $f = f_1$ . On the other hand, the phase of the spiral Fresnel zone plate is given by $\Phi_{SFZP}(f, L) = [\Phi_{FZP}(f) + L\theta]_{2\pi}$ , where $L$ is the topological charge, $f = f_2$ and $\theta$ is the azimuthal angle in the beam's cross section. The focal length values are selected such that $f_1 < f_2$ and the scatterer is positioned at the plane corresponding to $f_1$, as shown in Fig. 1(a). The two functions are multiplexed in the manner shown in Fig. 1(b). The resulting phase distributions at each step of the process are also shown there. $M_r(T)$ represents a virtual mask comprising a binary random pattern [0,1] synthesized with a transmittivity $T$, where $T$ is given by $N_1/(N_1+N_2)$. $N_1$ and $N_2$ are the number of pixels with value 1 and 0 respectively. As the virtual mask is applied directly to the spiral phase through the operation $\exp[j\Phi_{SFZP}]M_r(T)$ , $T$ can also be considered to control the fraction of the incident energy directed into the LG beam. A larger value of $N_1$, would increase the intensity of the LG beam with respect to the reference beam. Therefore, we will henceforth refer to $T$, as the splitting ratio.

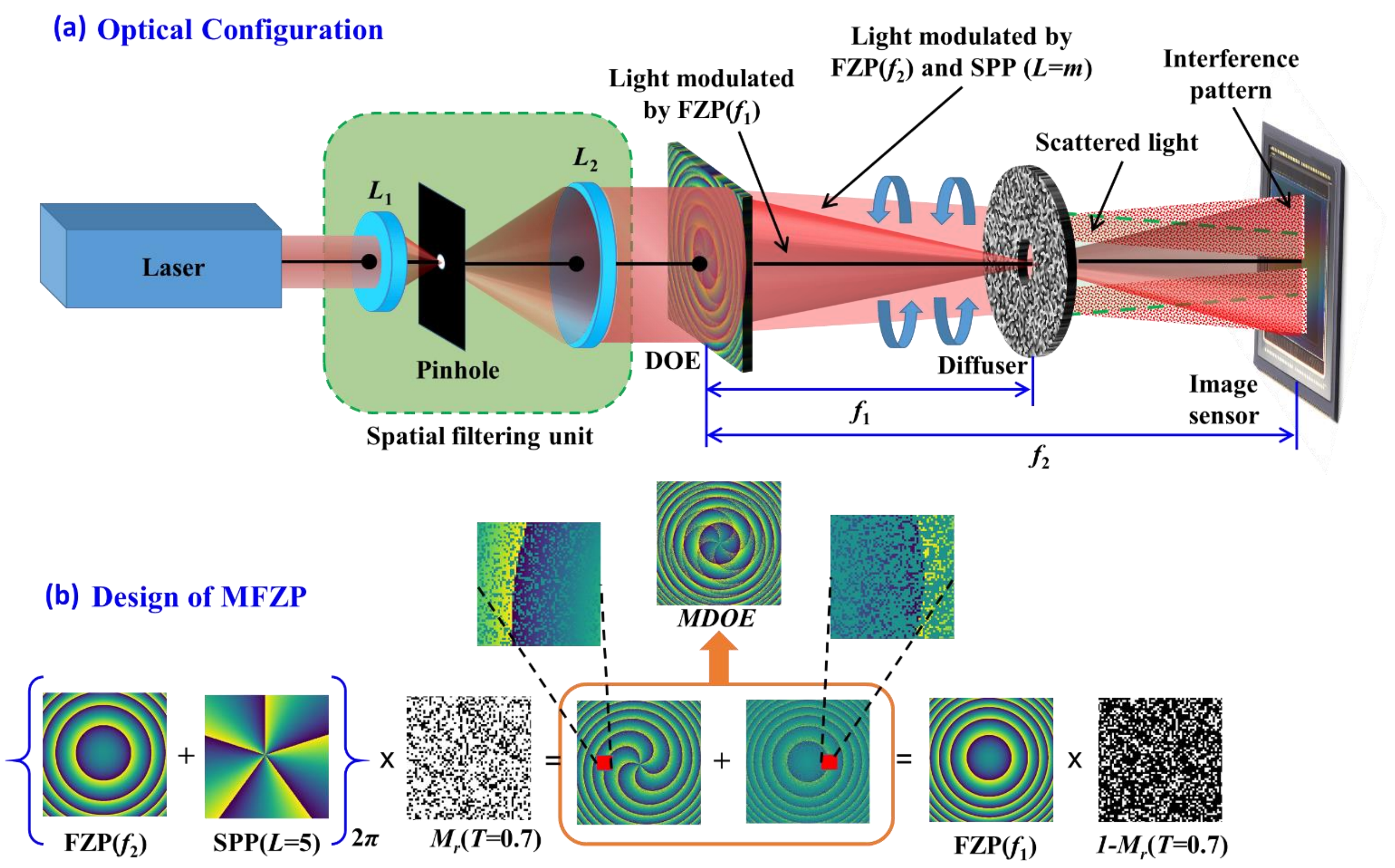

Fig. 1 (a) Optical configuration of the single channel, dual beam interferometer. The light from the laser is spatially filtered and collimated by a pinhole and refractive lenses $L_1$ and $L_2$. (b) Technique by which the MDOE is designed.

Collimated light from a coherent source is incident on the MDOE. This generates two co-propagating beams with complex amplitudes $exp[j\Phi_{FZP}]\{1 - M_r(T)\}$ and $\exp[j\Phi_{SFZP}]M_r(T)$. At the focal plane $f_1$, the former beam is focused to a spot, whereas the latter occupies a larger annular region. There is therefore, no overlap between the beams. On the other hand, at the focal plane $f_2$, the two beams spatially overlap resulting in an interference pattern, with intensity given by

$$I_1 = |\mathrm{E}_1 + E_2|^2, \tag{1}$$

where $E_1$ is a diverging spherical wavefront,

$$E_1 = \Im\left\{[1 - M_r(T)]\exp[j\Phi_{FZP}]\exp\left[j\frac{\pi R^2}{\lambda f_2}\right]\right\} \tag{2}$$

In Eqn. (2), $R$ is the radial distance from the axis. Neglecting complex constants results in the following expression for the spirally varying phase front $E_2$,

$$E_2 = \Im\{M_r(T)exp[-j\Phi_{SFZP}]\}. \quad (3)$$

In order to study the effect of scattering, a stack of annular scatterers, with phase $\Phi_S(x, y)$ in an annular region, is introduced with the center of the stack lying on the optical axis at focal plane $f_1$. The index $p$ is used to denote the number of layers in the stack. The resultant scattering phase therefore, is the modulo-2π phase addition of $p$ scattering layers (that form the stack). The intensity at the plane $f_2$ will be given by $I_2 = |\mathrm{E}_1 + E_2'|^2$, where the complex amplitude of the scattered vortex beam is

$$E_2' = \Im\left\{exp[j\phi_s]\Im\left\{M_r(T)exp[-j\Phi_{SFZP}]\exp\left[j\frac{\pi R^2}{\lambda f_1}\right]\right\}\exp\left[j\frac{\pi R^2}{\lambda(f_2-f_1)}\right]\right\}. \quad (4)$$

The degree of deterioration of the vortex beam can be measured by a cross-correlation between $I_1$ and $I_2$ in comparison to the autocorrelation of $I_1$. It is well-known that autocorrelation of a function gives the narrowest peak while the deviation from this ideal function can be quantified against the degree of deterioration.

## 3. Experiments

### 3.1 Fabrication of the MDOEs

MDOEs were designed as binary elements and therefore, the calculated function $\Phi_{MDOE}$ is binarized to have only two phase values [0, π] [35]. Since, binarization produces a zeroth diffraction order when there is a phase error during fabrication, an additional linear phase with an angle $\alpha$ = 0.03 radians was added to the MDOEs such that the required optical signal is moved away from the optical axis. The images of the design files for MDOEs with $T$ = 0.3, 0.5 and 0.7 for topological charges $L$ = 1 to 5 are shown in Fig. 2. The central area is magnified to clearly show the increase in the number of forks with a corresponding increase in the topological charge $L$. When $T$= 0.5, the incident energy is divided equally between both the diffractive functions, whereas for other $T$ values, one function of the MDOE will dominate changing the splitting ratio. The fabrication procedure is presented in Supplementary section S1.

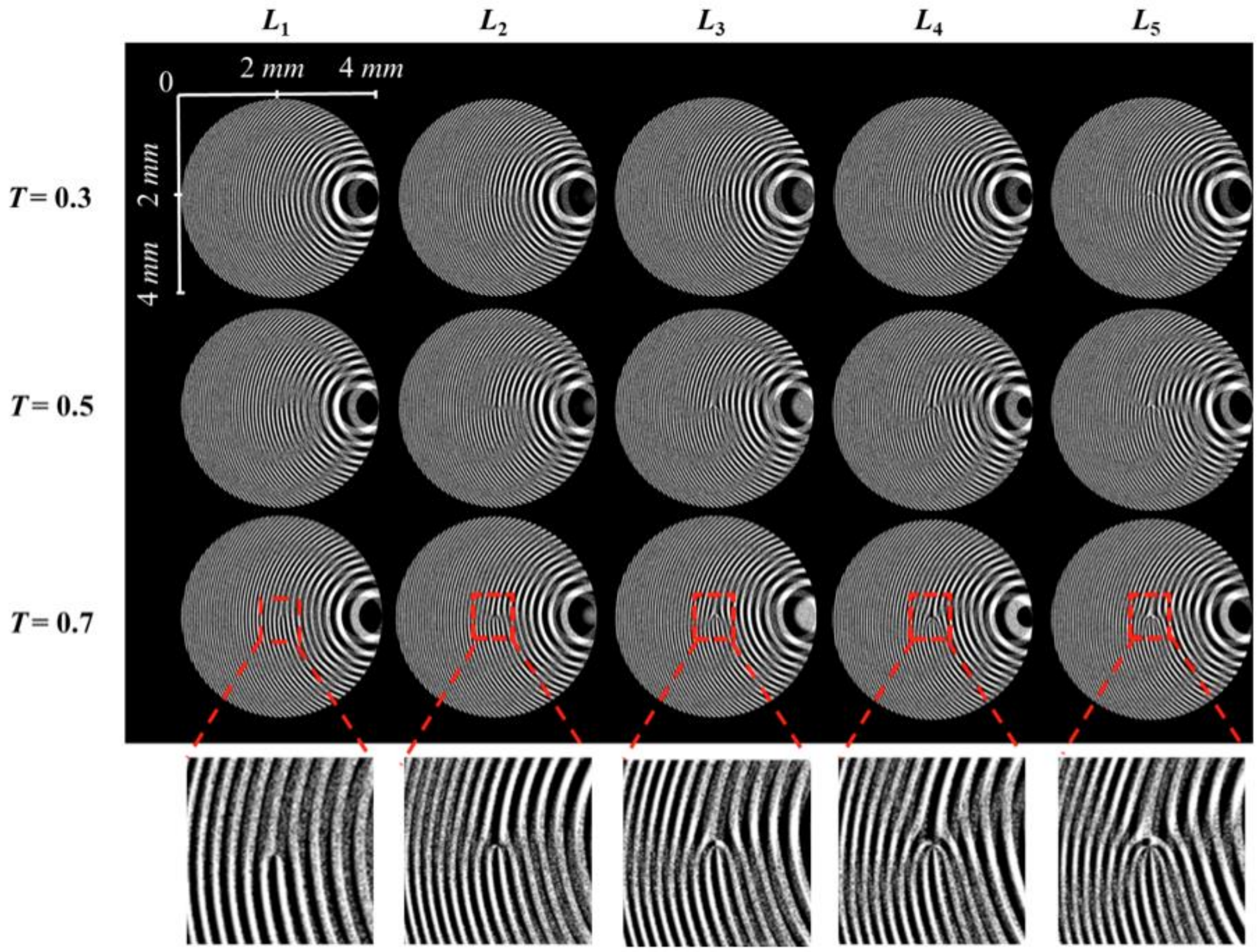

Fig. 2 Images of MDOEs for different topological charges $L$ = 1 to 5 and splitting ratios $T$ = 0.3, 0.5 and 0.7 designed with two-phase levels and a linear phase of $\alpha$ = 0.03 radians. The number of lines in the fork pattern increases with the topological charges.

### 3.2 Experiments

The experimental set up is shown in Fig. 3. Light from a He-Ne laser ($\lambda$=632.8 $nm$) is spatially filtered using a pinhole with a diameter of 100 $\mu m$ and a refractive biconvex lens $L_a$ ($f_a$ = 40 $mm$). The spatially filtered light is collimated by a second refractive biconvex lens $L_b$ ($f_b$ = 100 $mm$). The collimated light is incident on the MDOE, which generates several orders. Of interest are the spherical converging beam and a spiral-spherical beam, with focal lengths 25 $cm$ and 30 $cm$ respectively, generated in the +1$^{st}$ order. An image sensor (Thorlabs camera with 1024×768 pixels of size 4.65 $\mu m$) mounted at a distance of 30 $cm$ from the MDOE captures the interference of these two beams at this plane. The interference pattern will be between a spherical reference beam and either a scattered and or an unscattered LG beam, depending on whether or not the annular scatterer stack is introduced at a distance of $f_1$ (25 $cm$) from the MDOE. A study of the scattering characteristics of the scatterer stack is presented in Supplementary section S2. A neutral density filter (NDF) was used to reduce the light intensity and record the interference patterns without saturating the image sensor. The recorded intensity patterns in the two planes in the absence of a scatterer for different topological charges and splitting ratios are shown in Fig. 4.

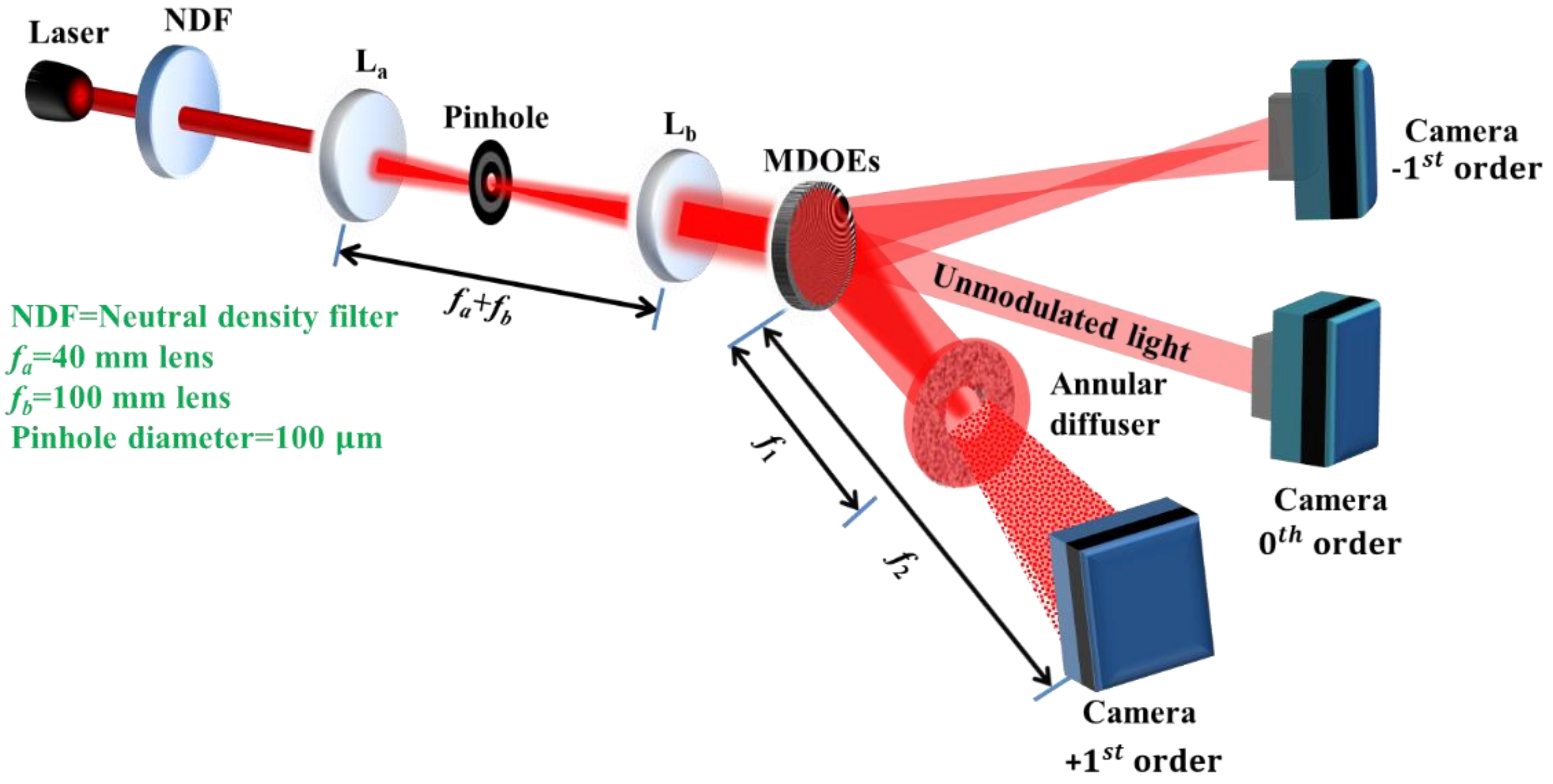

Fig. 3 Experimental setup for the evaluation of scatterers using the MDOE as a single channel interferometer.

Following this, the stack of scatterers is introduced in the plane $f_1$and the interference pattern is once again recorded at the plane $f_2$. The images of the recorded interference patterns for the different topological charges and splitting ratios are shown in the Fig. 5. Two sets of interference patterns are displayed in this figure, those taken with one or two scattering layers, corresponding to $p$ = 1 and $p$ = 2 respectively.

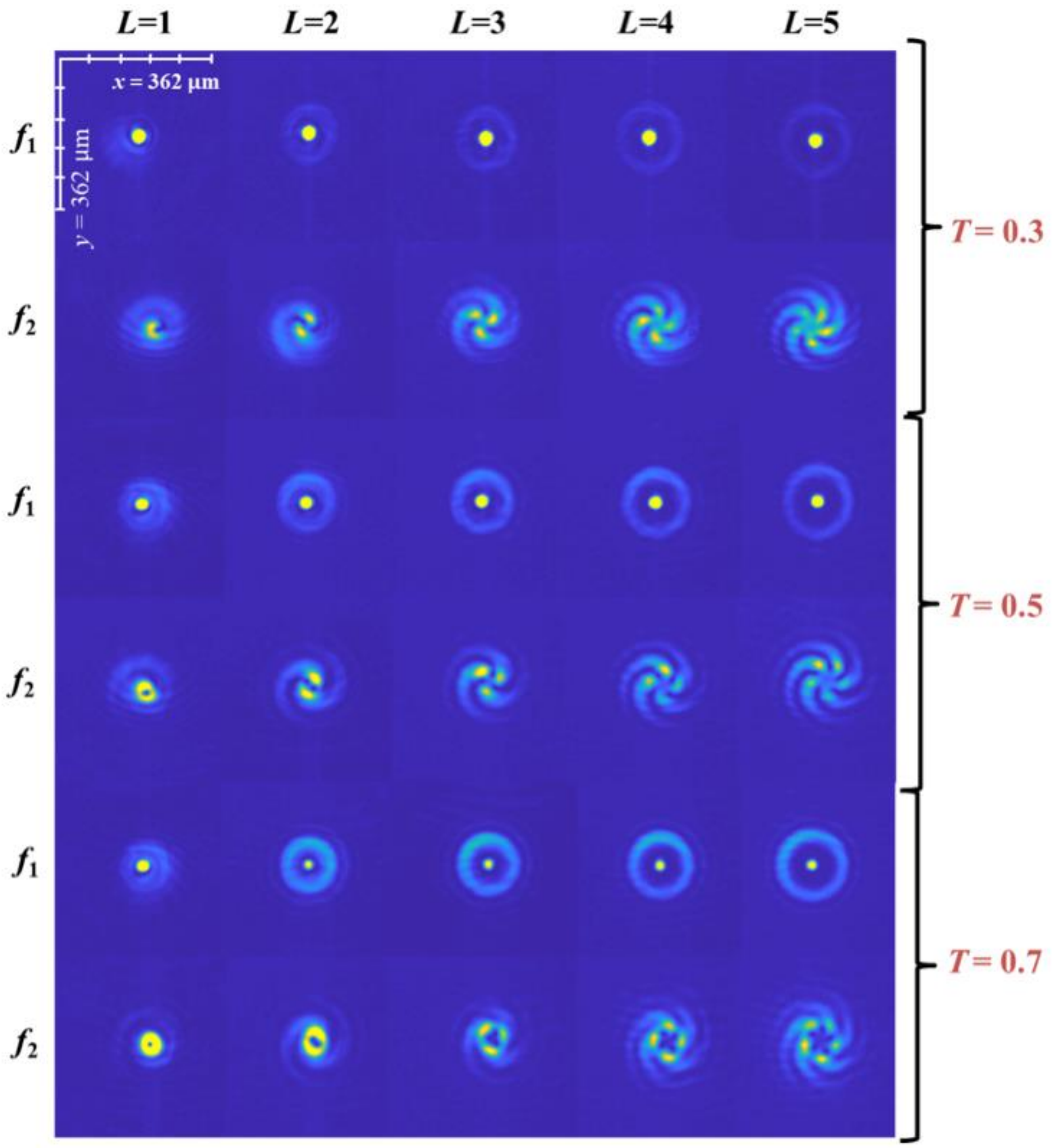

Fig. 4 Recorded intensity patterns without the scattering layers for different topological charges $L$=1 to 5, splitting ratios $T$ = 0.3, 0.5 and 0.7 and at two planes namely $f_1$ and $f_2$.

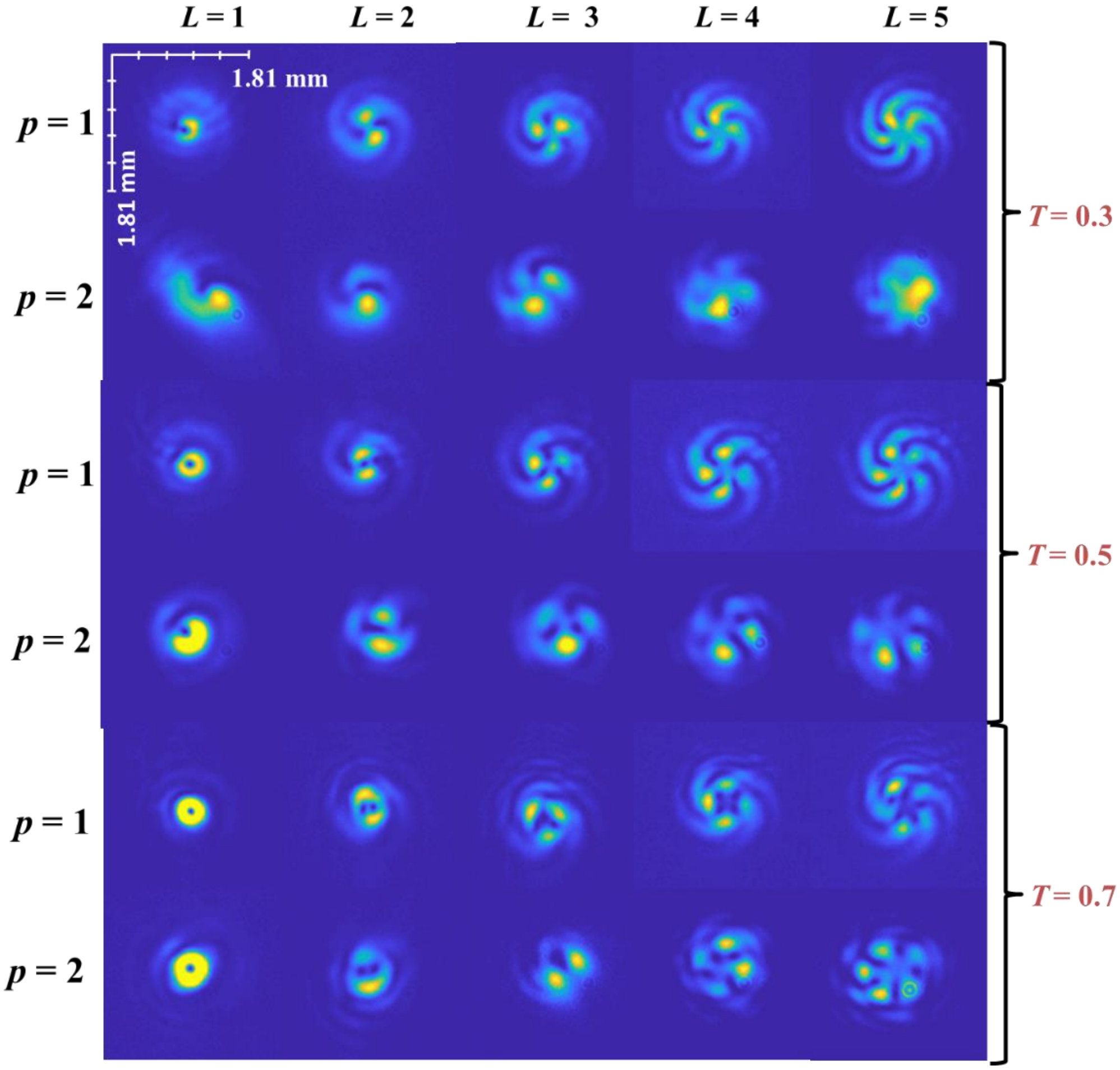

Fig. 5 Recorded intensity patterns at plane $f_2$ with the scatterer at $f_1$. The first rows present the results for one ($p$ = 1) and two ($p$ = 2) scatterers, whereas the columns are the results for the topological charges $L$=1 to 5. The experimental results for different splitting ratios ($T$ =0.3, 0.5 and 0.7) are presented in subsequent rows.

## 4. Discussion

The first point to note is that the resulting interference pattern contains a number of intensity lobes $m$ along the azimuthal direction that match the topological charge $L$. In the experiment, an LG beam of known charge is sent through the scatterer; any change in the charge of the beam would be easily picked from the number of lobes in the interference pattern.

From Fig. 4, it is seen that when the splitting ratio $T$ is increased, the relative intensity of the two beams in the plane $f_1$varied resulting in the ring pattern around the central spot becoming brighter. This is not surprising, as $T$ controls the amplitude of the LG beam, as described earlier. Figure 5 presents the interference patterns at plane $f_2$, with the scatterer in place. It is clear that the disturbance to the LG beam (that is the beam that sees the scatterer) causes the interference pattern to deviate from the case without the scatterer. The deviation is even more pronounced for the case two scattering layers ($p$ = 2). It is also observed that the deterioration to the interferogram is greater, when the LG beam's amplitude is weaker ($T$ = 0.3). The experimental results were verified by simulations, presented in Supplementary Section – 2.

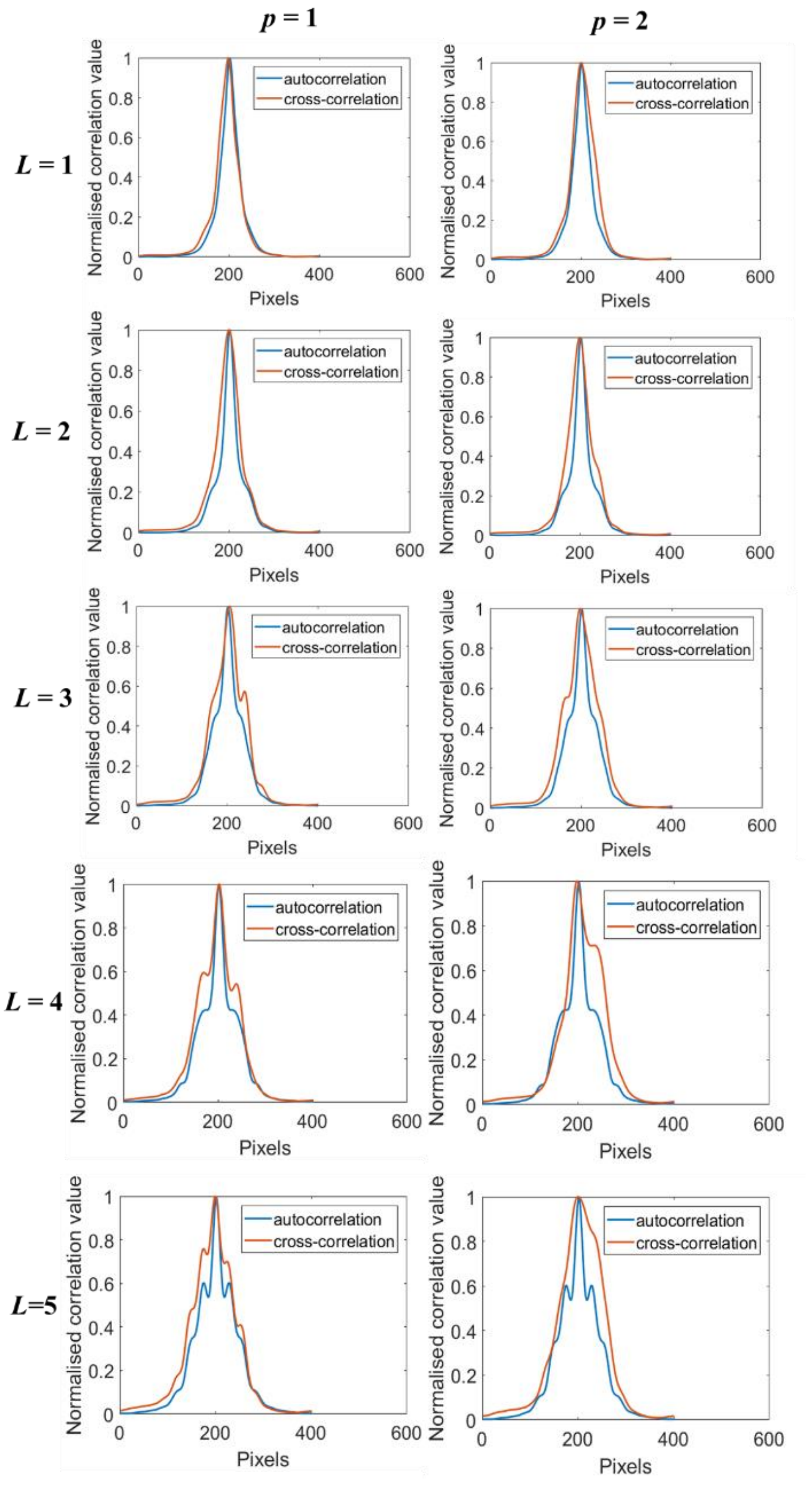

Fig. 6 Plot of the cross-correlation results for the scattering layers ($p$ = 1, $p$ = 2) for different topological charges $L$ = 1 to 5 and splitting ratios $M_r$ = 0.5.

In order to quantitatively compare the degree of deviation of the scattered case from the un-scattered one, the interference pattern recorded with the scatterer is cross-correlated with the interference pattern recorded without the scatterer. They are compared with the corresponding autocorrelation results of the un-scattered OAM beam. The cross-correlation results for $p$ = 1 and $p$ = 2 for $T$ = 0.5 are shown in Fig. 6. The difference between the $1/e^2$ radius of the cross-correlation and autocorrelation plots for topological charges $L$ = 1 to 5 are calculated and normalized with the corresponding autocorrelation radius. This is plotted in Fig. 7. The deviation of the cross-correlation function from the autocorrelation increases with the number of scattering layers. It is also found that for $p$ = 1 case, as the topological charge increases, the cross-correlation result has more deviation from its corresponding auto-correlation result. This says that lower charge OAM modes are less deteriorated by the scattering layer. It would be because we have used a dense scatterer, which is expected to deteriorate the stability of the propagating OAM mode strongly [36, 37], resulting in fracturing of the OAM mode. However in the $p$ = 2 case, the range of topological charge under study seems to be affected in the same way, which is why the deviation is more or less a constant with a slightly increasing trend.

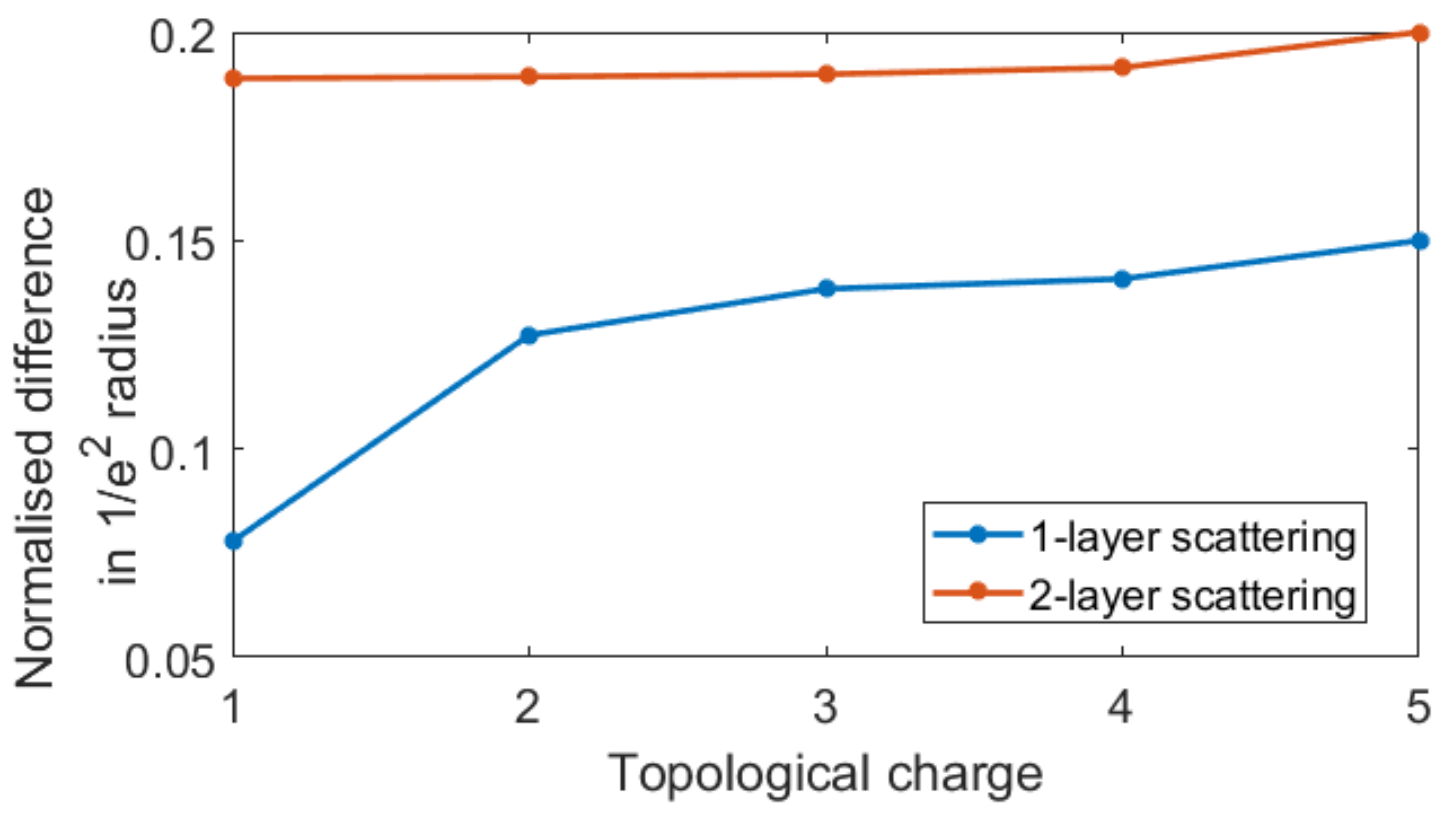

Fig. 7 Plot of the difference between the $1/e^2$ radius of the cross-correlation and autocorrelation plots for topological charges $L$ = 1 to 5 normalized with the corresponding autocorrelation radius, for the scattering layers ($p$ = 1, $p$ = 2) splitting ratio $T$=0.5.

## 5. Summary and conclusions

We have proposed and demonstrated a compact, single channel interferometer using a MDOE for the study of propagation of beams with orbital angular momentum through scattering layers. The MDOE generates two complex waves, where OAM is carried by only one of the waves and the other wave is used as a reference. The MDOE generated a focal plane suitable for modulating only the vortex, while transferring the reference wave without any modulation. The optical configuration was designed such that in an axial plane, the topological charge can be measured from the number of intensity peaks. In most existing optical configurations [25, 28], the study of beam propagation through scatterers involves bulky optical configurations with a minimum of two optical channels and many optical components. In the proposed method, the entire interferometer consists of only one optical element. The proposed single channel interferometer with a MDOE along with a blind cross-correlation method was found to measure the degree of deterioration of the beams carrying OAM. The experimental results matched with the results expected with the scatterers used. Clearly, the results would change dramatically according to the type of scattering media being studied. This conclusion is supported by the simulation results presented in the supplementary material. In particular, Supplementary Fig. 8 shows an improvement in the quality of the interference pattern (between reference and the scattered OAM beam) with an increase in the scattering ratio, for a constant phase retardation. However, Supplementary Fig. 9 indicates exactly the opposite behavior, as the phase retardation was increased along with the scattering ratio. More thorough studies and comparative analysis of different charged beams in scattering medium are needed. We believe that the proposed set-up and MDOE will allow exactly this. The efficiency of the demonstrated MDOE is limited due to the binary random multiplexing method and the efficiency can be increased by fabricating a greyscale version of the elements.

## Acknowledgements

The work was supported by the **LIGO R & D for India** project, run through the Inter-University Centre for Astronomy and Astrophysics, India. This study was done during a research stay of A.V at IIT Madras.

## Author Contributions

A.V. developed the idea and carried out the theoretical analysis and simulation for the research which was verified by S.B. A.V carried out the design of the MDOEs. The designs were fabricated and the experiment was carried out by S.J.L. The manuscript was written by A.V, S.B and S.J.L. All authors discussed the results and contributed to the manuscript. The entire research work was supervised by S.B.

## Additional Information

**Competing financial interests**: The authors declare that they have no competing interests.

# A compact single channel interferometer to study vortex beam propagation (Supplementary)

J. L. Sruthy,[1] A.Vijayakumar[1,2] and S. Bhattacharya[1]

[1]Department of Electrical Engineering, Indian Institute of Technology Madras, Chennai 600036, India.
[2]Centre for Micro-Photonics, Faculty of Science, Engineering and Technology, Swinburne University of Technology, Hawthorn VIC 3122, Australia.

## S. 1 Fabrication of MDOEs

The design patterns were transferred to chromium coated mask plates using laser fabrication method in a conventional mask writer. The images of the mask patterns are shown in supplementary figure S1. The amplitude masks were used for the fabrication of two level binary phase elements using UV lithography on borofloat glass ($t_g$=0.5 *mm*) coated with SU8-2002 NPR (MicroChem) with an index of refraction 1.58 for λ=632.8 *nm*. The resist layer was coated with a spin speed of 5000 *rpm*, acceleration of 500 *rpm*/*s* and time of 30 *s* followed by a prebaking at 95$^o$C for 1 *min*. The UV exposure was about 60 $mJ/cm^2$ followed by a post exposure baking at 95$^o$C for 2 *min*. The resist was developed by MicroChem's SU-8 developer for about 1 *min* and rinsed in IsoPropyl Alcohol and dried in nitrogen gas. The fabricated devices were hard baked at 120$^o$C for 30 *min*. The fabrication process is shown in figure S2. The microscope images of the fabricated elements are shown in the figure S3. The darker regions seen in the figure S1 and figure S3 are not due to underdevelopment of the resist but due to the spatial random pattern, which scatters light. A magnified version of the central area in the figure S3 shows that the dark regions are not actually underdeveloped.

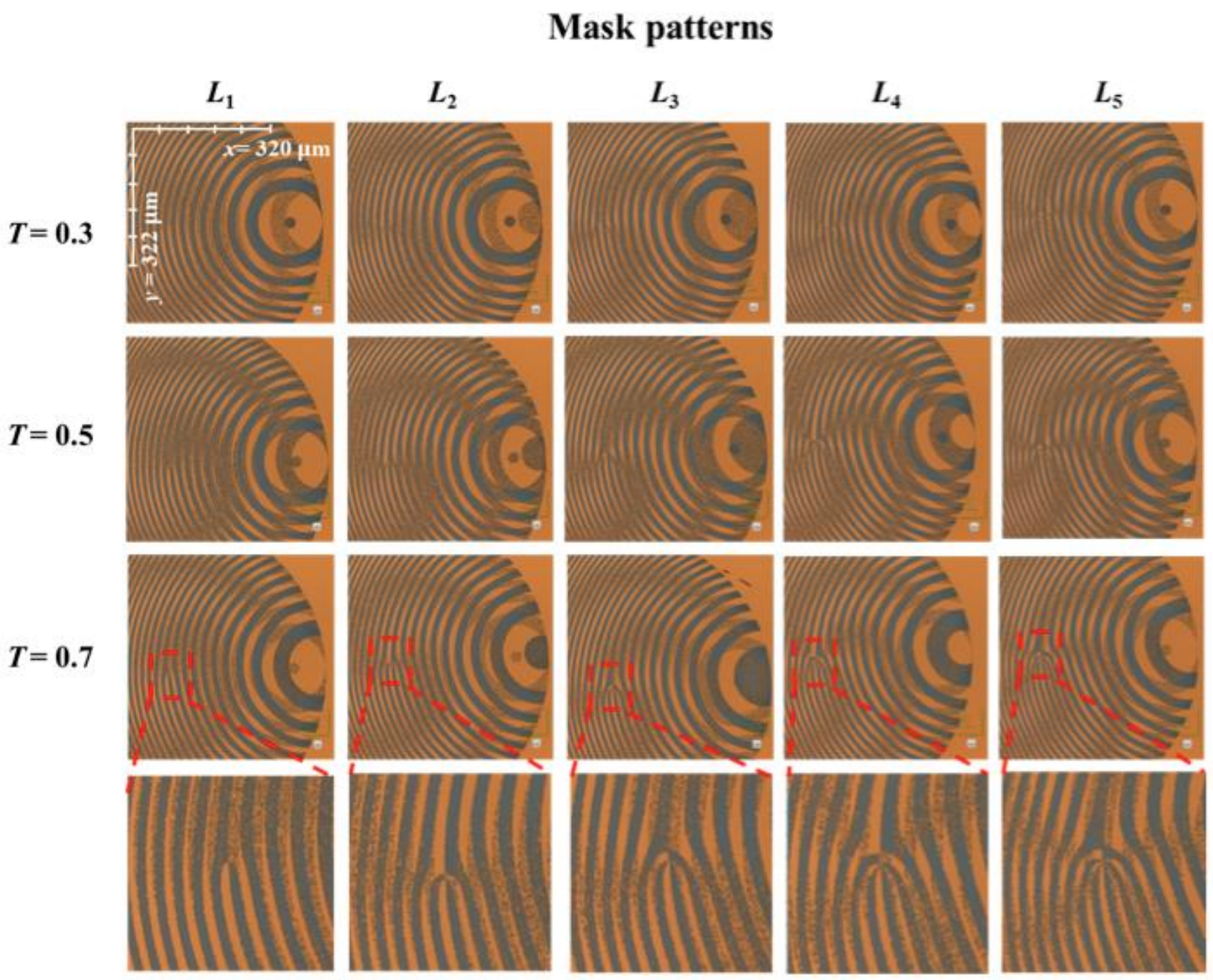

**Supplementary Figure S1:** Microscope images of the MDOE masks for different topological charges $L$=1 to 5, splitting ratios $T$=0.3, 0.5 and 0.7 and a linear phase with an angle $\alpha$=0.03 fabricated on chromium plates.

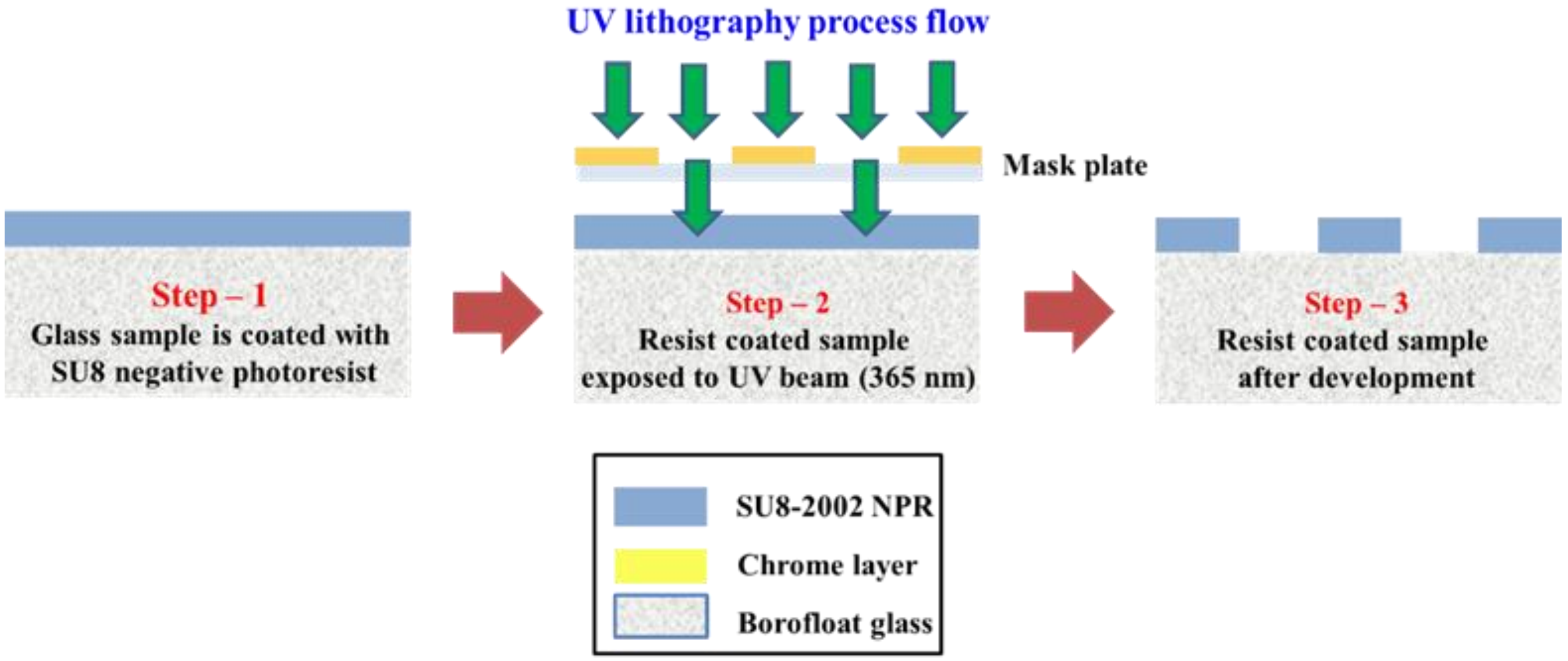

**Supplementary Figure S2:** Fabrication process for MDOEs using UV lithography.

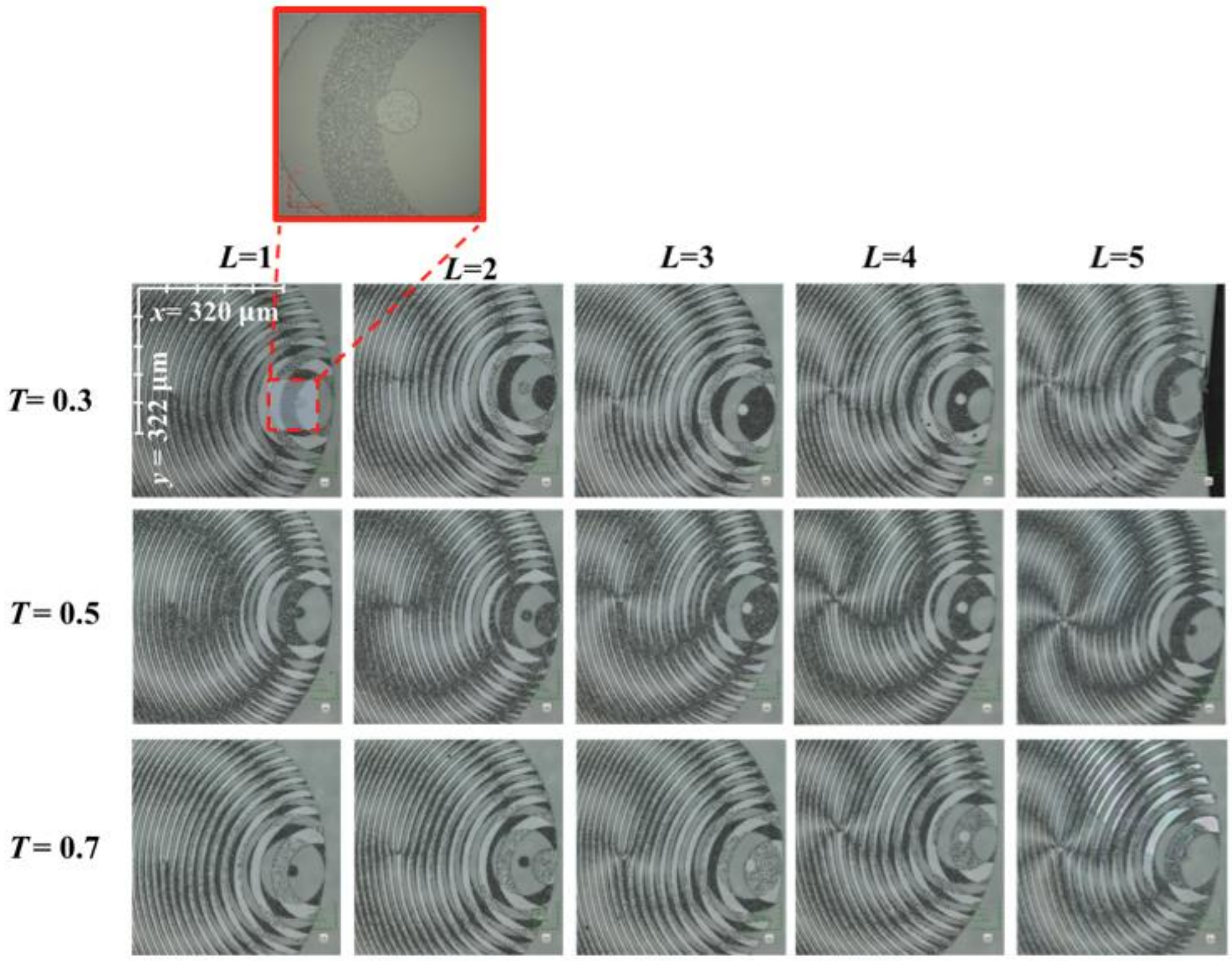

**Supplementary Figure S3:** Microscope images of the MDOEs for different topological charges $L$=1 to 5, splitting ratios $T$ = 0.3, 0.5 and 0.7 and a linear phase with an angle $\alpha$=0.03 fabricated on SU-8 resist layer.

## S.2 Simulation results

Simulation was done using Fresnel approximations at $\lambda$ = 632 *nm*, other parameters used were a sampling period of 4 $\mu m$, and a sampling space of 1000 × 1000 pixels, $\boldsymbol{f}(\boldsymbol{k_1})$ = 25 *cm* and $\boldsymbol{f}(\boldsymbol{k_2})$ = 30 *cm*. In the first step, the variation in the scattering ratio due to increasing number of layers in the stack of scatterers was studied. A scatterer is designed using Gerchberg Saxton algorithm (GSA) [1-3] with a scattering ratio given by $b/B$, where $B$ is the length of the spectrum domain and $b$ is the length outside which the intensity is zero. The process is shown in figure S4. A complex amplitude $C$ comprising a constant amplitude (White window at the input) and random phase distribution is Fourier transformed and the resulting amplitude is constrained to only have values within the scattering window of length $b$ while the phase is retained. The process is iterated to obtain the random phase distribution, which will have a scattering ratio of $\sigma = b/B$. The procedure is repeated with different initial random functions and a set of 5 weak scatterers are synthesized with negligible cross-correlation values with $b$ = 20 and $B$ = 1000. The cross-correlation between any two scatterers must be negligible so that when they are stacked, the effective scattering ratio increases. The images of the synthesized independent scatterers and the phase of the stack of weak scatterers when $p$ = 1-5 is shown in figures S5(a)-S5(e) and figures S5(f)-S5(j) respectively and their respective far-field diffraction patterns are shown in figures S5(k)-S5(o).

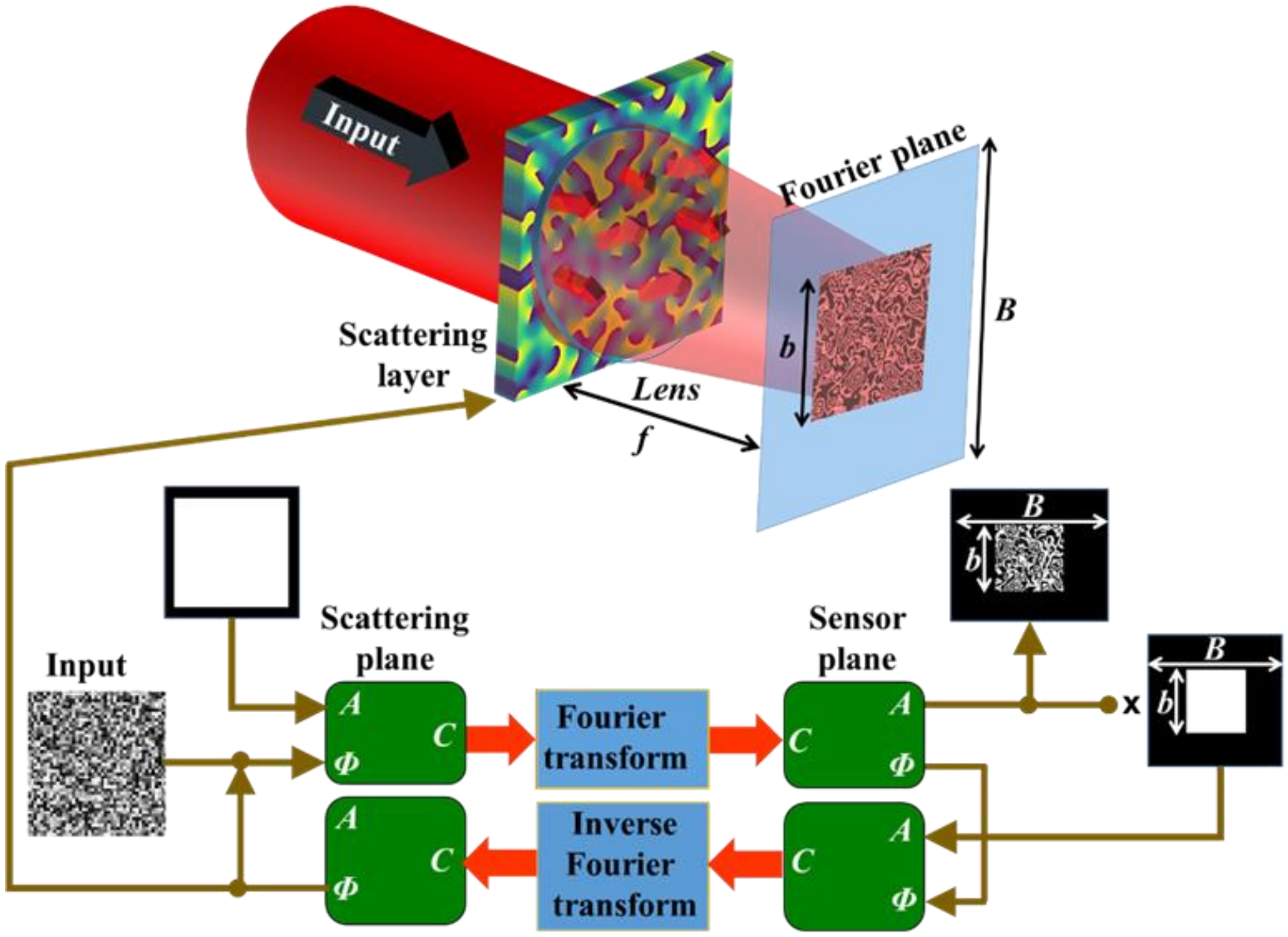

**Supplementary Figure S4:** Schematic of GSA and design of scattering mask with a scattering ratio of $b/B$; $C$-Complex amplitude; $A$ – Amplitude, $\Phi$ – Phase, $B$ – Length of the matrix and $b$ – length of the constraint.

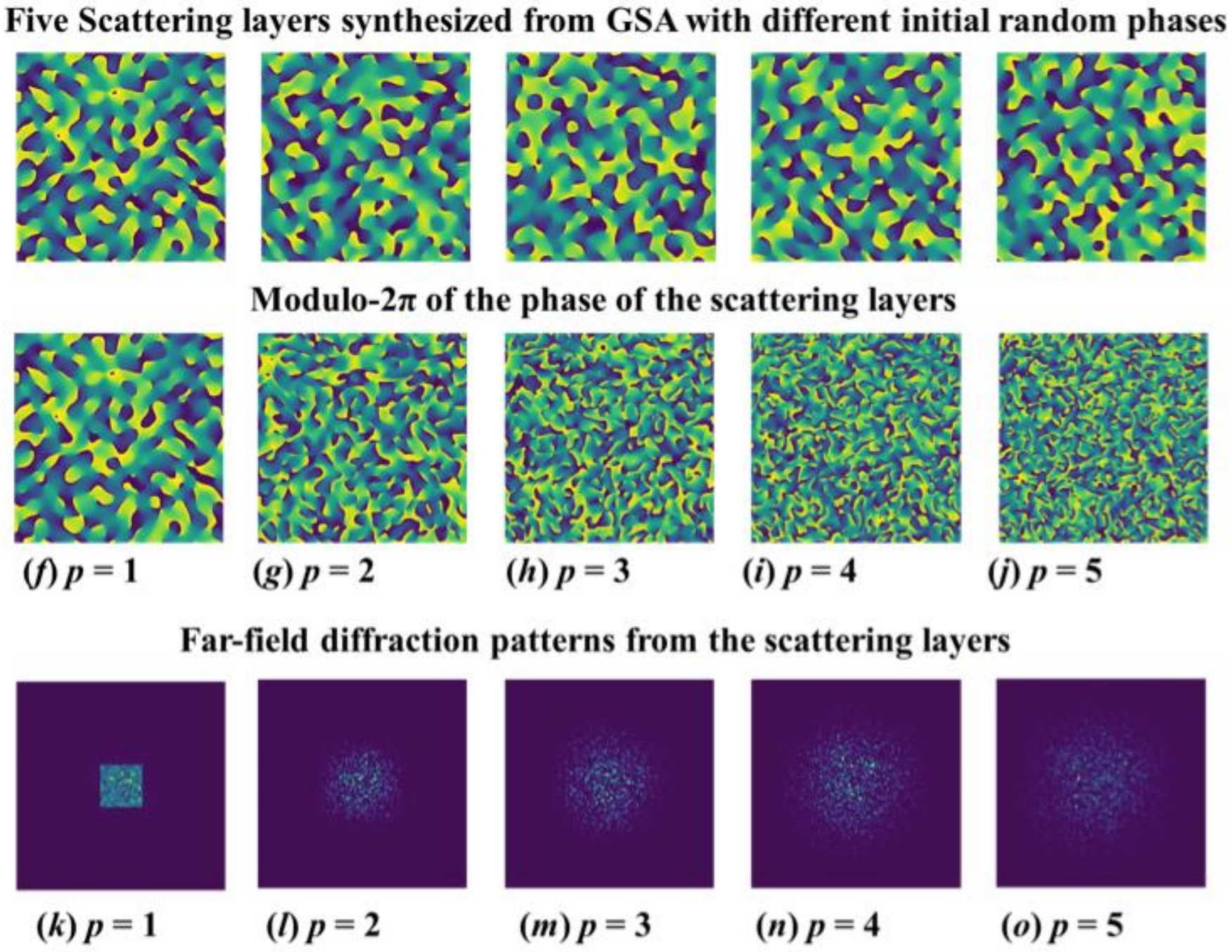

**Supplementary Figure S5:** Phase images of the scattering layers synthesized from GSA with a scattering ratio of *b*/*B*. Phase images of modulo-2π phase addition of multiple scattering layers for (a) $p = 1$, (b) $p = 2$, (c) $p = 3$, (d) $p = 4$, (e) $p = 5$.

Beam propagation is simulated through the different optical components in the absence of the stack of scatterers and the intensity patterns at $\boldsymbol{f}(\boldsymbol{k_1})$= 25 *cm* and $\boldsymbol{f}(\boldsymbol{k_2})$ = 30 *cm* for different topological charges $L$=1-5 are shown in the figure S6. All simulations were done at $T$ = 0.5, unless otherwise stated. The interference pattern at $\boldsymbol{f}(\boldsymbol{k_2})$ = 30 *cm* is stored as a library for implementation in the pattern recognition correlator [4]. Beam propagation is then simulated in the presence of a scatterer ($p$=1) capable of creating a maximum phase retardation of 0.2π – π in steps of 0.2π and the corresponding interference patterns for different topological charges are shown in figure S7. In this case, the scattering ratio remains constant but the maximum phase retardation is varied.

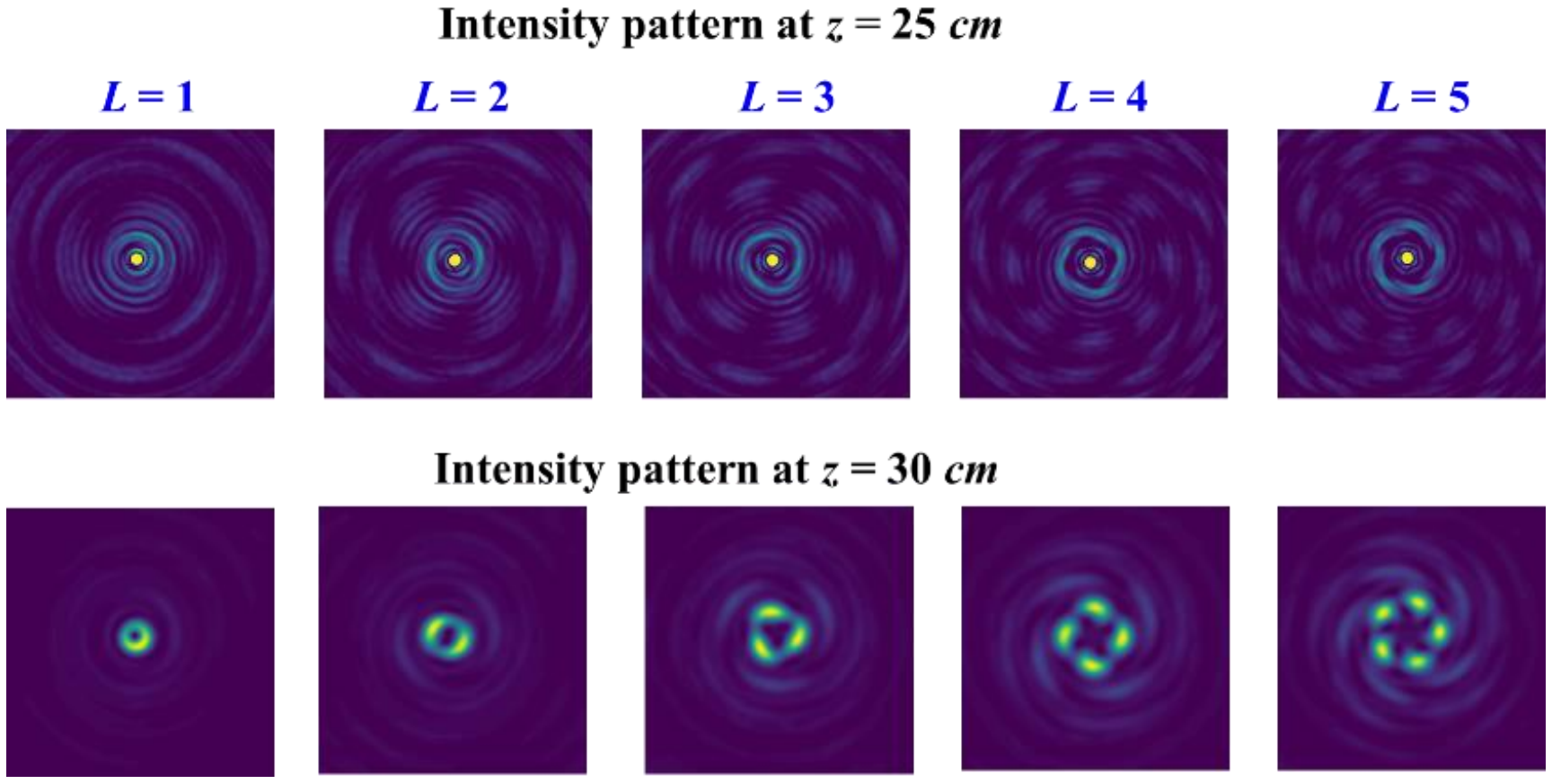

**Supplementary Figure S6:** Simulated intensity patterns at focal planes $\boldsymbol{f}(\boldsymbol{k_1})$ =25 *cm* and $\boldsymbol{f}(\boldsymbol{k_2})$ =30 *cm* for topological charges *L*=1 to 5.

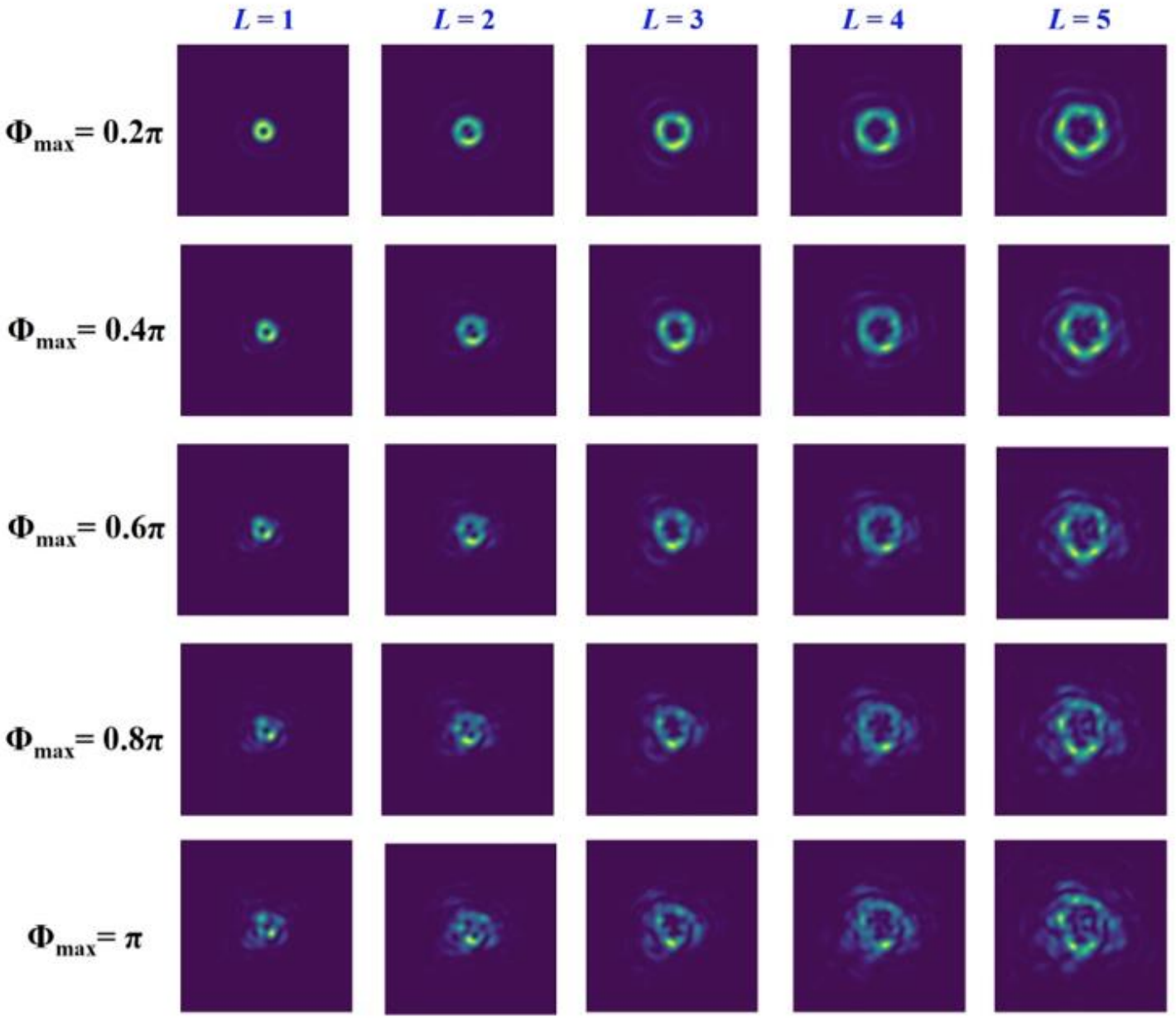

**Supplementary Figure S7**: Interference patterns for maximum phase retardation of 0.2π-π in steps of 0.2π and topological charges *L*=1 to 5.

The interference patterns for the case for different number of layers *p* = 1, 2 and 4 but for a maximum phase retardation of 0.6π in each case are shown in rows (1-3) of figure S8. In this case, the scattering ratio is varied but the maximum phase retardation was maintained constant and it is seen that with an increase in the scattering ratio, the results improve contrary to the belief that strong scatterers distort more compared to weaker counterparts. However, when the phase retardation was increased along with the scattering ratio, the behavior reversed. The interference patterns for the case of 0.6π phase retardation for *p* = 1, 2 and 4 with a maximum phase retardation of 0.6π, 1.2π and 1.8π are shown in figure S9. In this case both the phase retardation as well as the scattering ratio were increased.

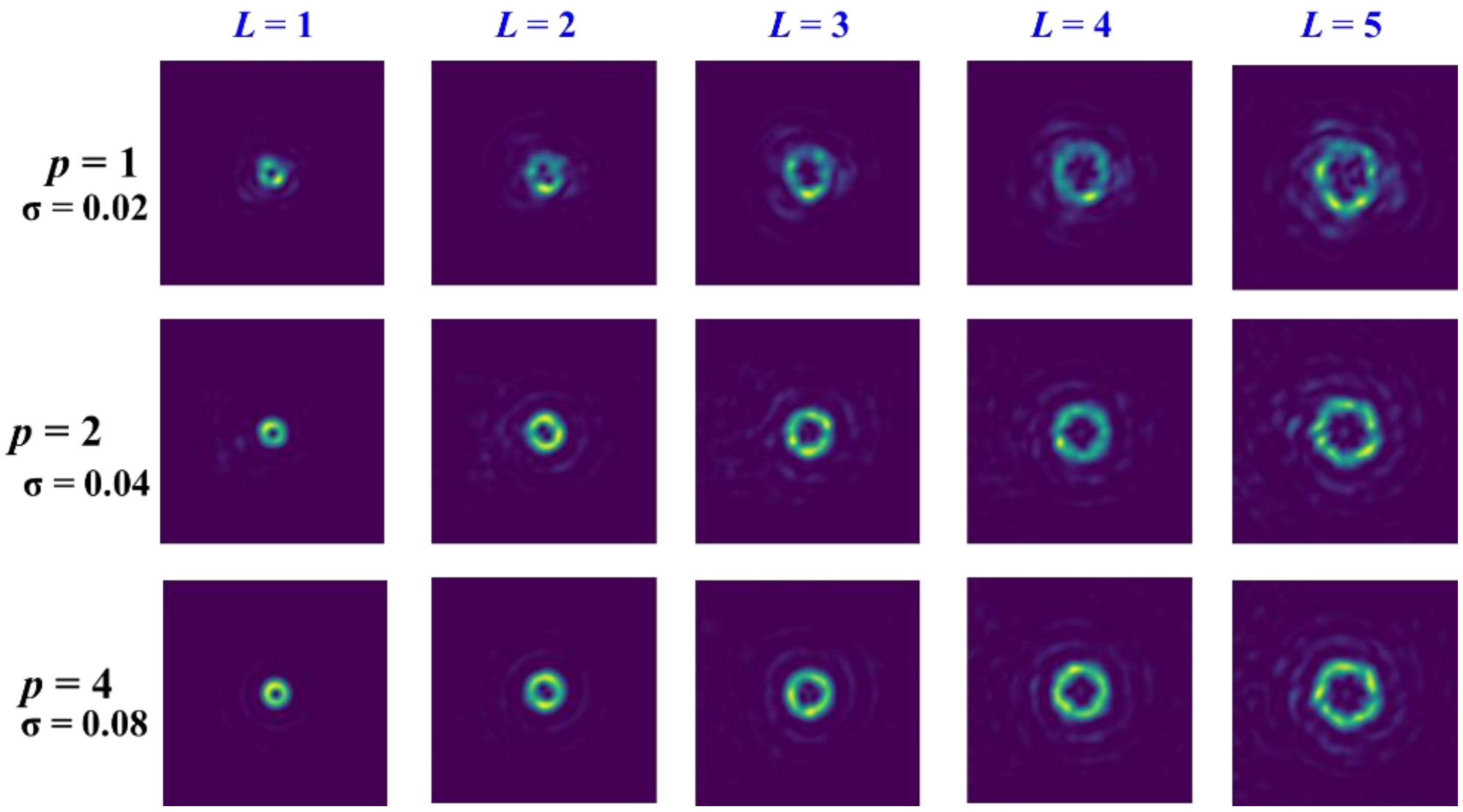

**Supplementary Figure S8:** Interference patterns for different scattering ratios and topological charges *L*=1 to 5.

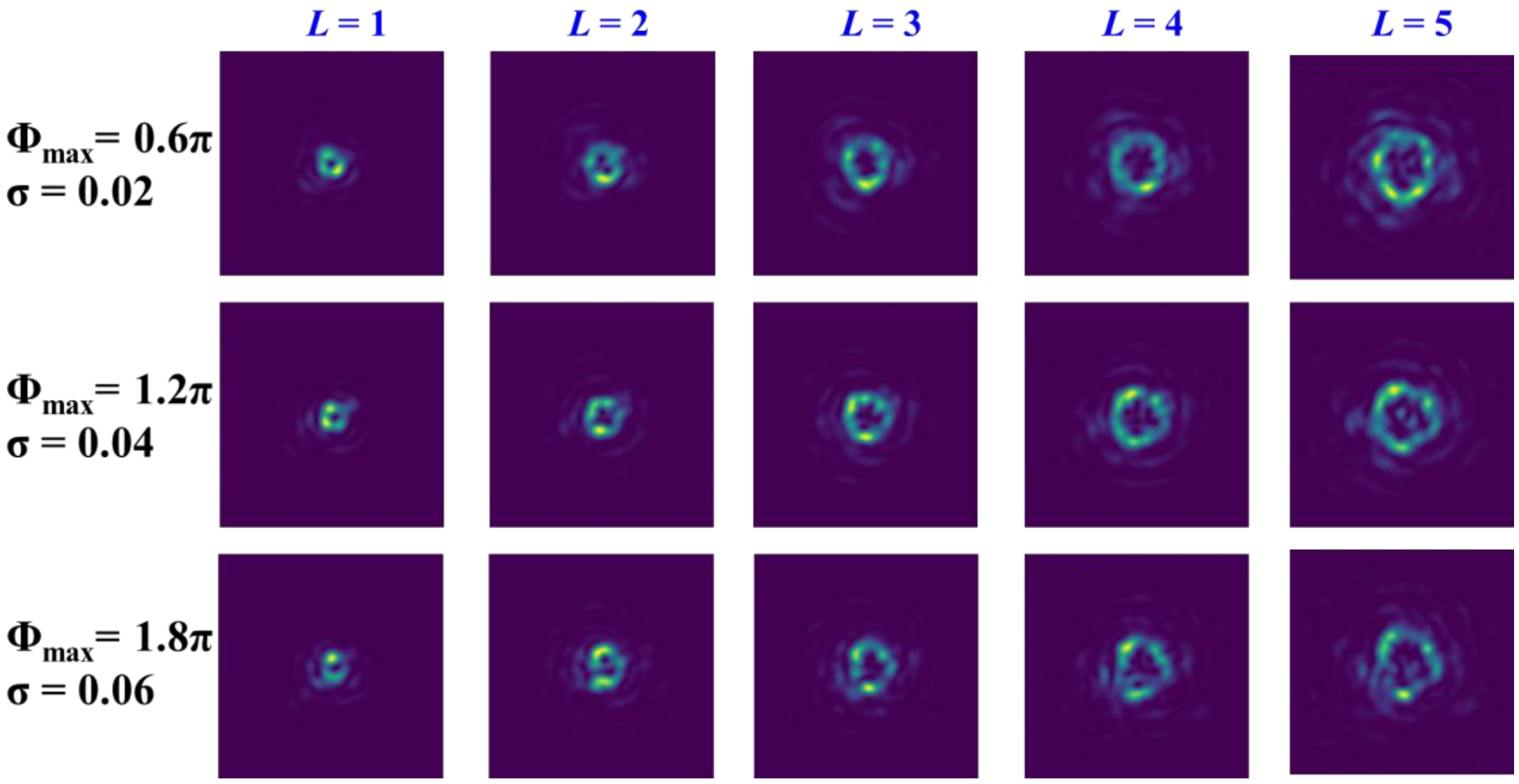

**Supplementary Figure S9:** Interference patterns for different number of layers and topological charges *L*=1 to 5.

The cross correlation results for a few of the above cases: $L$=3 for different phase retardation case with the respective elements from the library in the absence of the scattering layers are shown in figure S10. From the above observations, it can be seen that it is possible to measure the degree of deterioration of the vortex beam through thin scattering layers using a regular matched filter. The narrower the correlation function, better is the match between the two patterns. As seen from the Fig. 5, the broadening matches with the deterioration in the vortex signal. Therefore, the correlation technique can be used as a blind method to quantitatively measure the deterioration of the vortex signal.

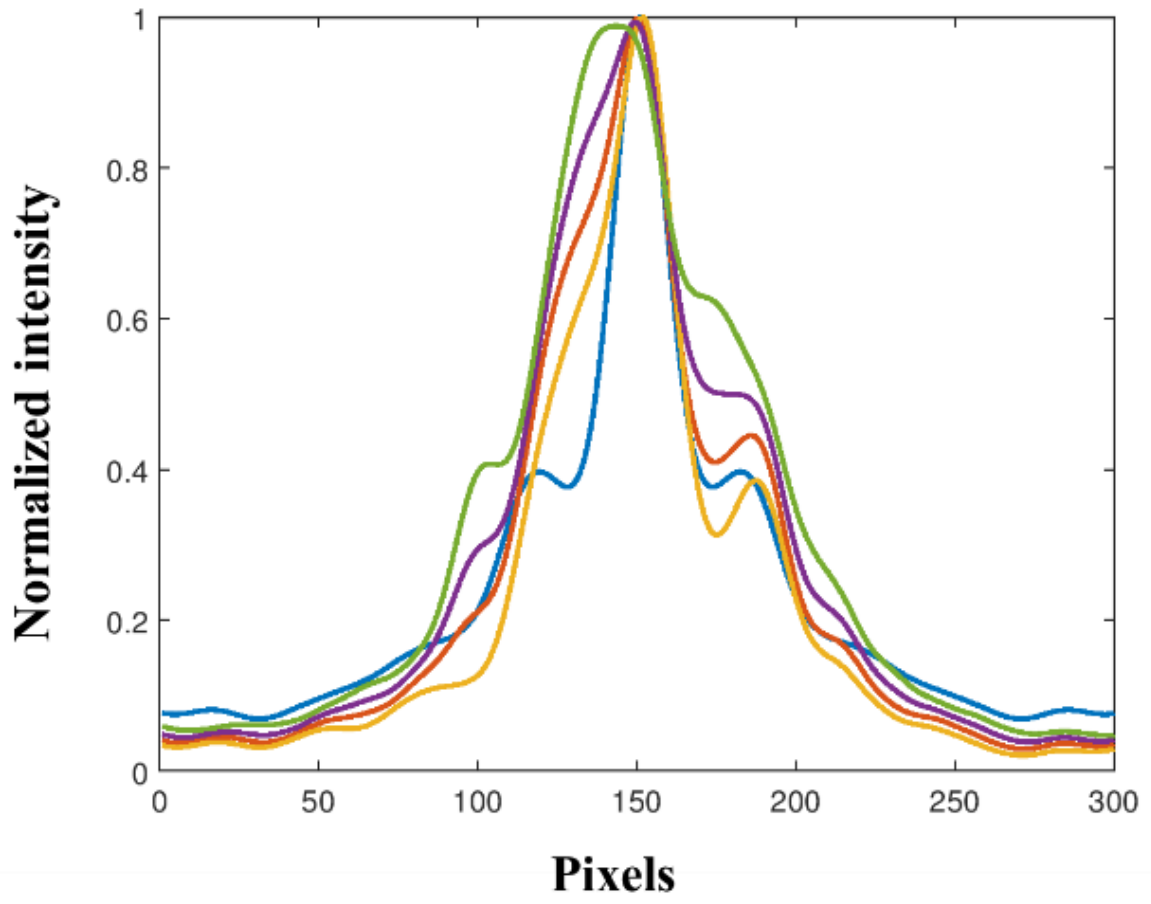

**Supplementary Figure S10:** Plot of the correlation results for $L$=3 between different phase retardations (0.2π-π in steps of 0.2π) and the case in the absence of a scatterer. Blue – 0.2π, Yellow – 0.4π, Meroon – 0.6π, Violet – 0.8π and Green – π.

### S.3 Study of scattering characteristics of scattering stack

The scattering characteristics of a stack of a scatterer was studied using an experimental setup as shown in figure S11. Light from a He-Ne laser ($\lambda$= 632.8 *nm*) is passed through a neutral density filter and a stack of scatterers. The stack of scatterers was created by stacking one scatterer over the other. The maximum scattering degree of the scatterer is measured using trigonometry as $\theta_{max}$= ($x$/2$L$). From the scattering degree, the maximum period of the scatterer can be approximated as $\Lambda_{max}$= (λ/sin $\theta_{max}$). The values of the scattering degree, period, etc., for the different number of scatterers that make the stack are given in Table – S1. With an increase in the number of layers, the scattering degree increased, while the effective scattering period decreased as described in the previous section.

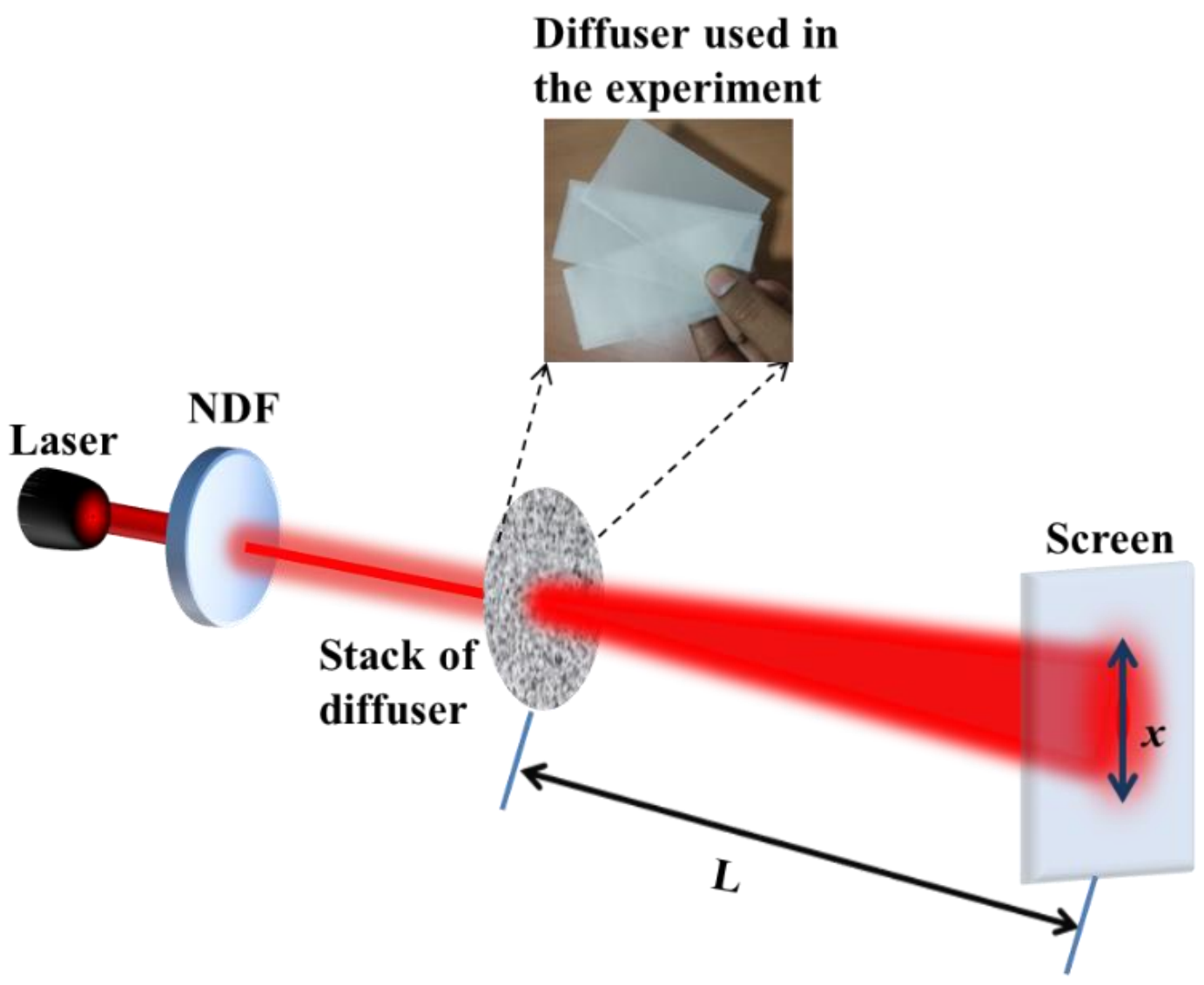

**Supplementary Figure S11**: Experimental setup for studying the scattering characteristics of a stack of scatterer.

**Supplementary Table S1. Scattering parameters for differnet number of scatterers**

| Number of Scattering layers | $x$ ($mm$) | $L$ ($mm$) | $\theta_{max}$ (degrees) | $\Lambda_{max}$ ($\mu m$) |
|---|---|---|---|---|
| 1 | 20 | 40 | 14 | 2.6 |
| 2 | 50 | 40 | 32 | 1.2 |
| 3 | 70 | 40 | 41.2 | 0.96 |
| 4 | 90 | 40 | 48.4 | 0.85 |